\documentclass[a4paper,fleqn,usenatbib]{mnras}

\usepackage{newtxtext,newtxmath}

\usepackage[T1]{fontenc}
\usepackage{ae,aecompl}

\usepackage{graphicx}	
\usepackage{amsmath}	
\usepackage{amssymb}	
\usepackage{siunitx} 
\usepackage{xcolor}
\graphicspath{{graphics/}}

\newcommand{\arecibo}{Arecibo Observatory}
\newcommand{\jbo}{Jodrell Bank Observatory}
\newcommand{\eff}{Effelsberg \num{100}-\SI{}{\m} radio telescope}
\newcommand{\deDCIV}{Bornim (Potsdam) LOFAR station}     
\newcommand{\Jiamusi}{Jiamusi 66-m telescope}      

\DeclareSIUnit\gauss{G}
\DeclareSIUnit\erg{erg}

\DeclareSIUnit\parsec{pc}
\DeclareSIUnit\lightyear{ly}
\DeclareSIUnit\year{yr}
\DeclareSIUnit\milliarcsecond{mas}
\DeclareSIUnit\lightsecond{lt-s}

\DeclareSIUnit\rydberg{Ry}
\DeclareSIUnit\magnitude{mag}
\DeclareSIUnit\jansky{Jy}
\DeclareSIUnit\h{$h$}
\DeclareSIUnit\hseven{$h$_7}

\DeclareSIUnit\hourra{\textsuperscript{h}}
\DeclareSIUnit\minutera{\textsuperscript{m}}
\DeclareSIUnit\secondra{\textsuperscript{s}}

\DeclareSIUnit\solarluminosity{\ensuremath{L_\odot}}
\DeclareSIUnit\solarmass{\ensuremath{M_{\odot}}}
\DeclareSIUnit\solarmassinenergy{\ensuremath{M_\odot|c^2}}
\DeclareSIUnit\solarradius{\ensuremath{R_\odot}}

\ExplSyntaxOn
\NewDocumentCommand {\SIeval} {omom}
  { \SI [#1] { \fp_eval:n {#2} } [#3] {#4} }
\ExplSyntaxOff

\usepackage{xspace}
\newcommand*{\eg}{e.g.\@\xspace}

\makeatletter
\newcommand*{\etc}{%
    \@ifnextchar{.}%
        {etc}%
        {etc.\@\xspace}%
}

\newcommand{\psr}{PSR~J0922$+$0638}

\newcommand{\mgr}{SGR~J1745$-$2900}

\title[Anomalous events of PSR~J0922+0638]{Multifrequency behaviour of the anomalous events of PSR~J0922+0638}

\author[G. Shaifullah et al.]{
G.~Shaifullah,$^{1,2,3}$\thanks{E-mail: shaifullah@astron.nl (GS)}
C.~Tiburzi,$^{3,2}$
S.~Os{\l}owski,$^{4}$
J.~P.~W.~Verbiest,$^{2,3}$
A.~Szary,$^{1,5}$
\newauthor
J.~K\"{u}nsem\"{o}ller,$^{2}$
A.~Horneffer,$^{3}$
J.~Anderson,$^{6}$
M.~Kramer,$^{3,8}$
D.~J.~Schwarz,$^{2}$
\newauthor
G.~Mann,$^{7}$
M.~Steinmetz,$^{7}$
and C.~Vocks$^{7}$
\\
$^{1}$ASTRON, The Netherlands Institute for Radio Astronomy, Postbus 2, {NL-7900} AA, Dwingeloo, The Netherlands\\
$^{2}$Fakult\"{a}t f\"{u}r Physik, Universit\"{a}t Bielefeld, Postfach 100131, 33501 Bielefeld, Germany\\
$^{3}$Max-Planck-Institut f\"{u}r Radioastronomie, Auf dem H\"{u}gel 69, 53121 Bonn, Germany \\
$^{4}$Centre for Astrophysics \& Supercomputing, Swinburne University of Technology, PO Box 218, H11, Hawthorn VIC 3122, Australia \\
$^{5}$Janusz Gil Institute of Astronomy, University of Zielona G\'{o}ra, Lubuska 2, 65-265 Zielona G\'{o}ra, Poland\\
$^{6}$Deutsches GeoForschungsZentrum, Telegrafenberg, 14473, Potsdam, Germany\\
$^{7}$Leibniz-Institut f\"{u}r Astrophysik Potsdam (AIP), An der Sternwarte 16, 14482 Potsdam, Germany\\
$^{8}$Jodrell Bank Centre for Astrophysics, University of Manchester, M13 9PL, UK
}

\date{Accepted XXX. Received YYY; in original form ZZZ}

\pubyear{2017}

\begin{document}
\label{firstpage}
\pagerange{\pageref{firstpage}--\pageref{lastpage}}
\maketitle

\begin{abstract}
PSR~J0922$+$0638 (B0919$+$06) shows unexplained anomalous variations in the on-pulse phase, where the pulse appears to episodically move to an earlier longitude for a few tens of rotations before reverting to the usual phase for approximately several hundred to more than a thousand rotations. These events, where the pulse moves in phase by up to \SI{5}{\degree}, have been previously detected in observations from $\sim$\SIrange{300}{2000}{\mega\hertz}. We present simultaneous observations from the \eff{} at \SI{1350}{\mega\hertz} and the Bornim (Potsdam) station of the LOw Frequency ARray at \SI{150}{\mega\hertz}. Our observations present the first evidence for an absence of the anomalous phase-shifting behaviour at \SI{150}{\mega\hertz}. Instead, the observed intensity at the usual pulse-phase typically decreases, often showing a pseudo-nulling feature corresponding to the times when phase shifts are observed at \SI{1350}{\mega\hertz}. The presence of weak emission at the usual pulse-phase supports the theory that these shifts may result from processes similar to the `profile-absorption' expected to operate for PSR~J0814$+$7429 (B0809$+$74). A possible mechanism for this could be intrinsic variations of the emission within the pulsar's beam combined with absorption by expanding shells of electrons in the line of sight. 

\end{abstract}

\begin{keywords}
pulsars: general -- pulsars: individual: J0922+0638, B0919+06 -- Stars, radiation mechanisms: non-thermal
\end{keywords}



\section{Introduction}\label{sec:Introduction}
Pulsars are rapidly rotating magnetised neutron stars which show extreme rotational stability on time scales of decades \citep{pt96, vbc+09}. However, almost all classical (or `slow') pulsars and a few millisecond pulsars (MSPs) exhibit significant and typically small-scale deviations from their stable rotational behaviour. These may include stochastic, wideband, pulse-to-pulse variations \citep{ovh+11, cordes93} due to unknown processes or, variations on short time scales like glitches \citep{rm69,rd69,mjs+16}, profile mode changes \citep{bac70a} and nulling \citep{bac70}. Apart from these, pulsars also display well known phenomena such as drifting subpulses \citep{bac70b}, polarisation modes \citep{gan97,st17}, microstructure \citep{jvkb01} and profile `absorption' where part of the profile is obscured, \citep[see \eg][]{rrs06}. A recent addition to these phenomena are `emission shifts' where the emission briefly shifts to an earlier longitude, first identified by \citet{rrw06} for PSRs~J1901$+$0716 (B1859$+$07) and J0922$+$0638 (B0919$+$06).

\psr{} is a bright, isolated pulsar which has a spin period of $\sim$\SI{0.43}{\s}. It is located about \SI[parse-numbers=false]{1.1^{+0.2}_{-0.1}}{\kilo\parsec} away \citep{ccl+01} and has associated dispersion and Faraday rotation measures of \SI{27.2986(5)}{\parsec\per\centi\m\cubed} and \SI{29.2(3)}{\radian\per\m\squared} \citep[DM and RM, values taken from][respectively]{srb+15,jhv+05}. Due to its relative brightness \psr{} has been observed at a number of frequencies and has been well-studied via polarimetric studies and timing analyses \citep[see \eg][]{scr+84, shabanova10, psw+15, wor+16}. 

The timing study of \citet{shabanova10} confirms a regular variation of the spin-down rate, $\dot{\nu} \equiv \Delta \nu /  \Delta T$, where $\Delta \nu$ is the change in rotational frequency of the pulsar and $\Delta T$ is the time in days. They find that the spin-down rate variation has a $\sim$\num{600} day periodicity and follows a rough sawtooth-like behaviour. They also detect a glitch on MJD~\num{55140} (5~November~2009) leading to a fractional decrease in the spin-period of $\delta$P$_{0}$/P$_{0}$ $\simeq$\num{7.69e-5}. 

As mentioned earlier, \psr{} also exhibits events marked by an aperiodic, rapid and continuous decrease in the observed emission longitude (or pulse-phase) by up to \SI{5}{\degree}. These were first identified using observations from the \arecibo{} by \citet{rrw06} at \SIlist{327;1420}{\mega\hertz}, who called them `emission shifts'. The emission shifts were also identified in observations with the \jbo{} by \citet{psw+15} who used pulsar timing analysis to measure variations in the spin-down rate of \psr{}. They find that the emission shifts (henceforth referred to as `events' for brevity) of \psr{} are not necessarily correlated with the two-state magnetospheric state switching cycle that \psr{} experiences. However, they suggest similar events occurring with a much smaller phase variation could be related to the observed state switching. Using the \Jiamusi{}, \citet{hhp+16} observed \psr{} over $\sim$\num{30} hours and were able to classify these events into four distinct morphologies. \citet{wor+16} searched for quasi-periodicities in these events using the datasets of \citet{rrw06}, complemented with new observations and archival data from the \arecibo{} as well as the observations of \citet{hhp+16} and found no evidence for such periodicities in the case of \psr{}. 
\section{Observations}\label{sec:Observations}
The observations presented in this article were carried out using the \eff{} and the Bornim (Potsdam) station of the LOw Frequency ARray \citep[LOFAR;][]{vwg+13}. At Effelsberg, the observations were carried out at \SI{1350}{\mega\hertz} using the central beam of the \num{21}-\si{\centi\metre} 7-beam receiver \citep{knm+06} and the PSRIX backend \citep{lkg+16} which records coherently dedispersed observations for up to \SI{300}{\mega\hertz} of bandwidth. Simultaneous observations using the \num{150}-\si{\mega\hertz} high band antenna (HBA) tiles of the \deDCIV{} (DE604) and the LOFAR und MPIfR Pulsare (LuMP) software\footnote{\url{https://deki.mpifr-bonn.mpg.de/Cooperations/LOFAR/Software/LuMP}} based backend were made possible by using the German Long Wavelength (GLOW)\footnote{GLOW is the German Long-Wavelength consortium, an association of German universities and research institutes who share an interest in using the radio spectral window at meter wavelengths for astrophysical research. The International LOFAR stations in Germany, described in \citet{vwg+13}, are owned and run by GLOW member institutions and more details about the GLOW network are available here: \url{https://www.glowconsortium.de/index.php/en/}. We use the acronym to identify the operating mode, as well as the scheduling and monitoring software that allow the use of the GLOW stations for the observations presented here.} mode. The observation details are presented in \autoref{Tab:obsdetails}. For both the observing systems, data were coherently dedispersed and folded modulo the spin period to produce \num{10}-\si{\second} subintegrations and recorded with a final phase resolution of \SI{0.35}{\degree} per phase bin.

\begin{table}
\centering                                                                                                                                                                                                                 
\caption{Details of observations carried out with the \deDCIV{} and \eff{} in November, 2017.}                                                                                                                                                                
\label{Tab:obsdetails}                                  
\begin{tabular}{@{}l
S[table-auto-round,table-format=-2.2,table-column-width = 1.6 cm,table-number-alignment = left]
S[table-auto-round,table-format=-2.2,table-column-width = 1.6 cm,table-number-alignment = left]
}
\hline
{Telescope}&\multicolumn{1}{c}{Effelsberg}&\multicolumn{1}{c}{DE604}\\
\hline
Source name					&\multicolumn{2}{c}{J0922+0638}			\\
Receiver name					&{P217-3}		&{HBA}			\\
Name of the backend instrument			&{PSRIX}		&{LuMP}			\\
Centre frequency (\si{\mega\hertz})		&1347.5			&153.80859375		\\
Weighted freq. (after RFI removal)		&1352.983		&152.115		\\
Bandwidth (\si{\mega\hertz})			&200			&71.484375		\\
Dispersion measure (\si{\parsec\per\cm\cubed})	&27.2951(6)	        &27.2951(6)       	\\
Number of pulse phase bins			&1024			&1024			\\
Number of frequency channels			&128			&366			\\
Polarisations recorded				&{Full Stokes}		&{Full Stokes}		\\
Sub-integration duration (\si{\s})		&10.00	                &10.00                  \\		        
Number of sub-integrations			&703			&703		        \\
Modified Julian Date				&57705			&57705			\\
\hline						
\end{tabular}					 
\end{table}

The observations were post-processed using the \texttt{PSRchive} software suite\footnote{\url{http://sourceforge.psrchive.org}} \citep{hvm04,sdo12}, radio frequency interference (RFI) was removed using the `median zapping' option of the \texttt{paz} tool from \texttt{PSRchive}, followed by selective reweighting of strongly affected channels and subintegrations using a script from the \texttt{COASTGUARD} 
data processing pipeline \citep{lkg+16}. The timing solutions used in the `folding' of the archives were corrected for spin-period offsets, measured using the \texttt{Tempo2} pulsar timing package \citep{hem06}. The DM used for both the observations was measured from the LOFAR observations, by splitting them into four bands and fitting for the DM while the remaining timing parameters were held fixed. The data were polarisation calibrated using the \texttt{pac} tool from \texttt{PSRchive}. The method used for the Effelsberg utilises standard, position offset observations with a calibrated noise diode while for the GLOW data, we use the method outlined in \citet{nsk+15}. 
\section{Discussion}
\begin{figure}
\def\svgwidth{\linewidth}{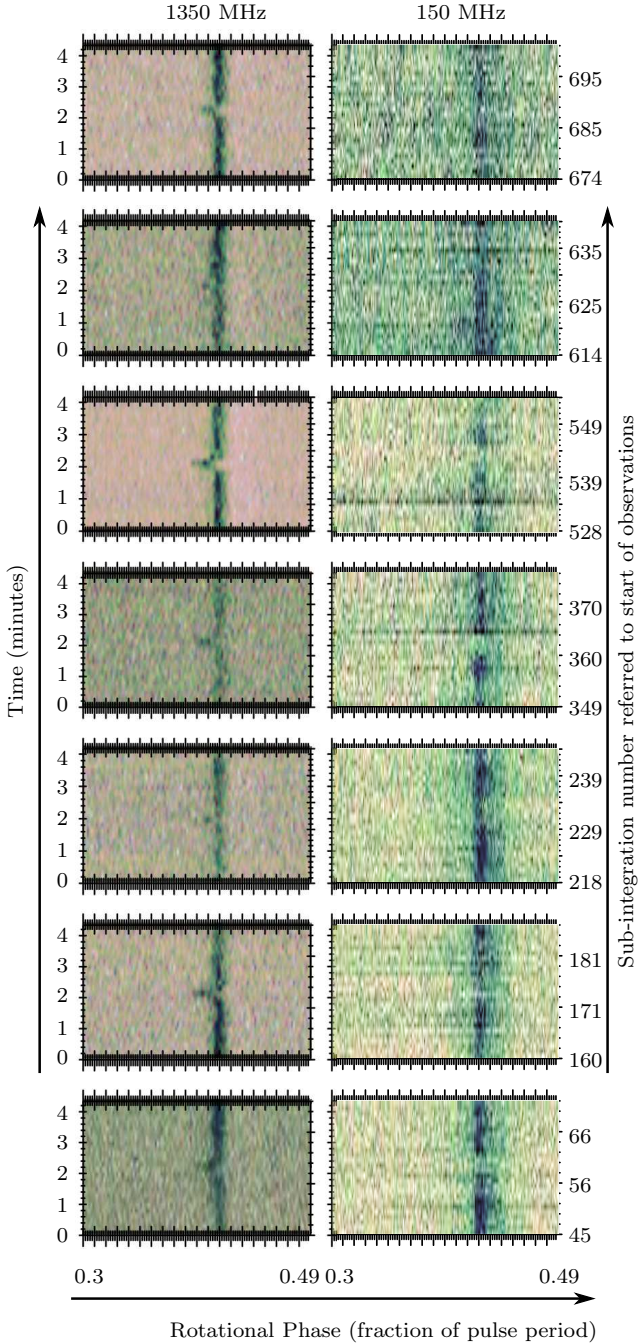}
\caption{Observed intensity colour-mapped from lighter to darker shades plotted against the observation time (or subintegration number) and the rotational phase for the simultaneous observations from Effelsberg (left) and DE604 (right). 
\label{Fig:time_plot_combined}}
\end{figure}

Plots showing the intensity (colour-mapped to darker shades for higher values) as a function of the observing time and the rotational phase are presented in \autoref{Fig:time_plot_combined}. On the left are the \SI{1350}{\mega\hertz} Effelsberg observations while the \SI{150}{\mega\hertz} LOFAR observations are plotted on the right. 

\begin{figure}
\def\svgwidth{\linewidth}{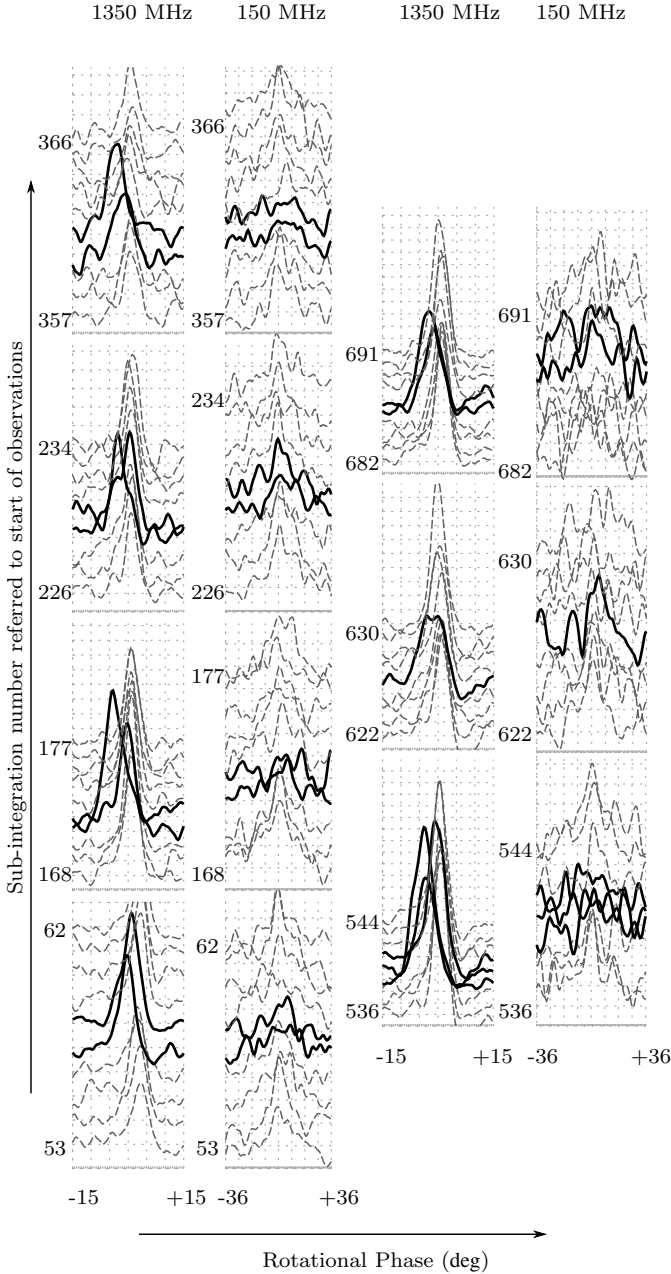}
\caption{Subintegration plots of sections of the data presented in \autoref{Fig:time_plot_combined}. Each curve shows the integrated flux for a \num{10}-\si{\s} subintegration. The amplitudes of the integrated pulse profiles have been scaled independently for the \SIlist{1350;150}{\mega\hertz} data.  The bold, solid lines show the `event' subintegrations. Note the absence of a distinct profile for the corresponding regions of the \SI{150}{\mega\hertz} plots.
\label{Fig:subints}}
\end{figure}

Using the \SI{1350}{\mega\hertz} observations to detect the events, we find a total of seven events, as shown in \autoref{Fig:time_plot_combined}, five of which are easily visible and a further two which are only identified in the subintegration plots of \autoref{Fig:subints}. While it is not possible to recover the differing morphologies of these events as presented in the detailed single-pulse analysis by \citet{hhp+16} we note that all seven events show differing degrees of pulse-phase variation and duration in our observations whose finest time resolution is limited to \num{10}-\si{s} long subintegrations. 

\begin{figure}
\def\svgwidth{\linewidth}{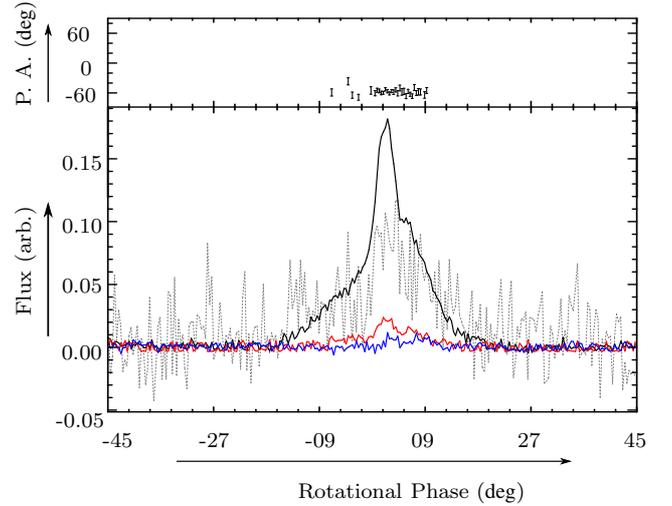}
\caption{Integrated profile for the LOFAR observations with the `event' subintegrations removed (solid black line) and only those subintegrations which are associated with an emission shift event added together separately (dotted gray line). The red and blue curves show the linearly and circularly polarised intensities, i.e., the $\left|\mathrm{L}\right|\equiv\left|\mathrm{Q + iU}\right|$ and $\left|\mathrm{V}\right|$ Stokes components respectively. The top panel shows the change in the position angle measured with a significance greater than 2$\sigma$.
\label{Fig:LOFARprofile}}
\end{figure}

For each event, we also present subintegration plots in \autoref{Fig:subints}. The \SIlist{1350;150}{\mega\hertz} plots have been scaled and smoothed independently. In spite of the scaling and smoothing, the low signal-to-noise ratio (S/N) due to excess RFI at the LOFAR bands often degrades the visibility of the integrated pulse profile. To facilitate comparison,  \autoref{Fig:LOFARprofile} shows the integrated total intensity profile (black curve) for all the subintegrations which do not correspond to the events. The red and blue curves show the integrated linearly and circularly polarised intensities, respectively, while the top panel shows PA measurements with a significance of 3$\sigma$ or more for the same subintegrations. The dotted, gray curve shows the integrated total intensity profile for the event subintegrations, which have been plotted with thick black lines in \autoref{Fig:subints}. 

The majority of previous work on the emission shifts of \psr{} were compiled using observations at \SIlist[range-units=single,list-final-separator = { or }]{327;1420;2400}{\mega\hertz} \citep[see eg.][]{wor+16,rrw06,psw+15,hhp+16} where these events are marked by a gradual shift of the emission phase to an earlier longitude without any change in the average flux density, as can be seen in the \SI{1350}{\mega\hertz} observations in \autoref{Fig:time_plot_combined}. However, at \SI{150}{\mega\hertz} the events display a marked absence of emission at any shifted phase along with a significant decrease in the observed intensity at the usual phase. The integrated pulse profile for the events (dotted gray curve in \autoref{Fig:LOFARprofile}) shows a significant but decreased amount of emission, although the profile shape is distorted due to the low S/N as only fourteen subintegrations were combined to create the dotted gray profile compared to the remaining \num{689} for the solid black profile. Some of the observed power in the dotted gray profile is also due to the fact that the pulsar takes only a few to about thirty revolutions for the emission to diminish from or be restored to the typical observed flux density levels \citep[see \eg][]{hhp+16,rrw06}, which can occur on a timescale much shorter than the \num{10}-\si{\second} subintegrations in our data. As a result, a few of the subintegrations contain a fractional amount of power from the usual emission state.

In most cases the diminished emission at lower frequencies precedes and outlasts the emission shift at higher frequencies by a few subintegrations or tens of rotations. The change from `normal' to diminished emission at the lower frequencies has a gradual onset and recovery, similar in the smoothness of its variation to the emission shift at higher frequencies although on distinctly longer timescales. However, in keeping with its non-conformal behaviour, these diminished emission phases of \psr{} show distinct variations amongst themselves as well. For example, in \autoref{Fig:subints}, for the coupled event at subintegrations \numrange{226}{234}, the decrease in emission lasts for less than a couple of subintegrations while for the event at subintegrations \numrange{357}{366}, the decrease precedes the shift by at least two subintegrations and does not recover to the normal profile for at least one subintegration after the shift has ended.

It is evident that the integrated data presented in our observations can only offer broad insights into this phenomenon and to gain a full description of the physical processes driving these changes, high S/N single-pulse data are necessary. However, even without the single-pulse data, it is possible to test some of the proposed origins of these events. 
\begin{itemize}
\item `Absorption' \citep{rrw06}~: The phenomenon of `absorption' where the emission region is partly obscured has been one of the most promising candidates for explaining the complicated nature of the emission shifts of \psr{}. The archetypical example of this phenomenon is PSR~J0814$+$7429 (B0809$+$74) \citep{rrs06} where part of the inherently broad profile is absorbed or occulted leading to distinct portions of the profile not being detected at certain frequencies. 

We follow the approach presented in \citet{rrw06}, where this mechanism is first applied to explain these emission shifts and it is made clear that `absorption' implies any physical process that might lead to part of the profile not being visible. However, in order to explain our observations this `absorption' must also have a time dependent efficiency, along with the expected frequency dependent behaviour. This is an added complexity over the case of PSR~J0814$+$7429. There is a further constraint on the absorbing medium that it must always lead to an absorbed portion of profile at the leading edge at the LOFAR band while it must be more efficient at the regular emission phase only during the emission shift events at high-frequencies. It is therefore abundantly clear that even if `absorption' is the cause for the observed behaviour, the physical process leading to `absorption' in this case might easily be very different from the typical examples.  

\item Binary companions in light-cylinder orbits \citep{wor+16}~: 
The
absence of a true periodicity in the emission shifts of \psr{}
as well the extreme orbits and compositions necessary for such
companions make such objects unlikely in the first place. Moreover,
the necessity of similarity in the emission shifts at high and low
frequencies is clearly in contrast to the observations.

\item Differentially-corotating (entire) magnetosphere~:
\citet{ym17} propose that the observed emission from a pulsar is due to the combination of the entire magnetosphere differentially-corotating with respect to the stellar surface, structures around the magnetic pole \citep[similar to the carousel model of ][]{dr99} and a moving visible emission region. They used seven components to model the observed profile of \psr{}, and this appears to predict the observed emission shift at \SI{1400}{\mega\hertz}. As the components move due to variations in the magnetospheric rotation rate, this leads to different portions of the components crossing the line of sight, leading to the observed shift. While results for a lower frequency are not presented, the authors do point out that their approximations include ignoring a possible dependence of the magnetospheric plasma rotation rate on the radial distance from the centre of the pulsar. If such a dependence exists, it might be possible to explain the observed pseudo-nulling features at \SI{100}{\mega\hertz} although this would still necessitate a very fortunate arrangement of the line of sight.

However, the change in the magnetospheric properties in the \citet{ym17} model enters via their unphysical `$y$' parameter. It is difficult to reconcile the variation of this global parameter with the absence of any correlation between the events and the magnetospheric state switching of \psr{} observed by \citet{psw+15}.
 
\item Quasi-stable magnetosphere~: There are hints that in some pulsars nulling and mode changing may be manifestations of the same phenomenon \citep[see e.g.][]{wmj07}. It was suggested by \citet{tim10} that magnetospheres can operate in different quasi-stable states with different sizes of the open field line zone or different current distributions in the radio emission region, or both. A magnetosphere can occasionally switch between these states and depending on our line of sight, we see either a different part of the emission cone, or we miss the entire cone resulting in the ``null'' state. It was also shown that modest variations in the beam size can be accompanied by large variations in the pulsar spin down rate. The combination of the effects of different current density distributions and different sizes of the emission cone could explain our observations of PSR J0922+0638, however as pointed out by \citet{psw+15} there is no correlation between the events and the magnetospheric state switching.

\item Frequency selective nulling~: Using simultaenous multifrequency observations from \SIrange{300}{4850}{\mega\hertz}, \citet{bgk+07} show that PSR~J1136$+$1551 (B1133$+$16) shows `frequency selective' nulling where nulls appear at one frequency even when emission is observed at another. 
  The observations presented here might be explained by invoking frequency selective nulling only at the LOFAR bands, combined with either `absorption' or apparent motion of the line of sight with respect to the emission regions. However, it should be noted that in the case of PSR~J1136$+$1551, the frequency selective nulling is not imited to the lowest frequencies only. 
  
\item Intrinsic variations in the pulsar's beam combined with free-free absorption along the line of sight~: \citet{lrk+15} suggest that for the \mgr{} (also known as the Sgr~A$^{*}$ magnetar or the Galactic centre magnetar), expanding electronic ejecta due to outbursts can explain rapid variations in the observed radio spectrum of the magnetar. A similar mechanism in combination with intrinsic variations of the emission in the pulsar's beam would also be capable of explaining the observed phase shifting at higher frequencies and decreased emission in the LOFAR band. 

\end{itemize}
\section{Conclusions}
In summary, simultaneous observations of \psr{} using the \eff{} and the \deDCIV{} demonstrate that during the previously identified phase shifts of the high-frequency emission, the emission at lower frequencies is strongly diminished, often appearing as null-like features accompanied by a marked absence of emission at any shifted phase. These null-like features strongly disfavour hypothetical binary companions. While models involving differential magnetospheric rotation rates could explain the observed variations, those models appear to be in tension with previous work which find no correlation between the switching of the magnetospheric states \psr{} is believed to exhibit. It is possible that profile `absorption' is indeed the mechanism due to which this emission shift is observed. However, the models of local `absorption' in the emission region currently do not address the levels of complexity required to explain the time-varying yet distinct behaviour above and below $\sim$\SI{250}{\mega\hertz}. These events might also arise due to a combination of intrinsic variations in the pulsar's beam with expanding screens of electrons along the line of sight.

However, a detailed analysis is beyond the scope of this letter and interested colleagues wishing to inspect the data presented here are encouraged to contact Jun.Prof. Verbiest or any of the leading authors for access to these and other, similar datasets.  
\section*{Acknowledgements}
The authors are grateful to the staff at the \eff{} and the GLOW consortium for their continued support for the respective telescopes which makes observations such as those presented here possible. GS acknowledges support from the Netherlands Organisation for Scientific Research (NWO; TOP2.614.001.602), and is grateful to Gemma Janssen, Cees Bassa and Yogesh Maan for useful discussions. SO acknowledges Australian Research Council grant Laureate Fellowship FL150100148. A. Sz. received funding from the NWO under project ``CleanMachine'' (614.001.301). JK received funding from the German Federal Ministry of Education and Research (BMBF) under grant 05A14PBA (Verbundprojekt D-LOFAR III).

This paper is based (in part) on data obtained with LOFAR equipment. LOFAR \citep{vwg+13} is the LOw Frequency ARray, designed and constructed by ASTRON. 
We acknowledge the support and operation of the GLOW network, computing and
storage facilities by the Forschungszentrum J\"{u}lich, the Max-Planck-Institut f\"{u}r Radioastronomie and Bielefeld
University and financial support by the German states of Nordrhein-Westfalia
and Hamburg.
 



\bibliographystyle{mnras}
\bibliography{J0922+0638}






\bsp	
\label{lastpage}
\end{document}